\documentclass[aps,prl,twocolumn,amsmath,amssymb,nofootinbib,superscriptaddress,longbibliography]{revtex4-1}

\usepackage{times}
\usepackage[pdftex]{graphicx}
\usepackage{dcolumn}
\usepackage{bm}
\usepackage{amsmath}
\usepackage{indentfirst}
\usepackage{float}
\usepackage[colorlinks]{hyperref}
\usepackage[dvipsnames]{xcolor}
\usepackage{xcolor}
\usepackage{subfigure}
\usepackage{algorithm}
\usepackage{algorithmicx}
\usepackage{algpseudocode}
\usepackage{amsmath}
\usepackage{verbatim}
\usepackage{makecell}
\usepackage{braket}
\usepackage{multirow}
\usepackage{lineno}

\definecolor{citecolor}{HTML}{2E3092}
\hypersetup{colorlinks=true,allcolors=citecolor}
\let\oldcite\cite
\renewcommand{\cite}[1]{\textcolor{citecolor}{\oldcite{#1}}}
\begin{document}
\title{Demonstration of a Qubit Control Scheme Compatible with Superconducting Digital Signals}
\author{Chenyin Sun}
\thanks{These authors contributed equally to this work.}
\affiliation{\adda}
\affiliation{\addb}
\affiliation{\addc}
\author{Yu Xu}
\thanks{These authors contributed equally to this work.}
\thanks{xuyu0922@ustc.edu.cn}
\affiliation{\adda}
\affiliation{\addb}
\affiliation{\addc}
\author{Zihua Chen}
\thanks{These authors contributed equally to this work.}
\affiliation{\adda}
\affiliation{\addb}
\affiliation{\addc}
\author{Dayu Li}
\affiliation{\adda}
\affiliation{\addb}
\affiliation{\addc}

\author{Tianzuo Sun}
\affiliation{\adda}
\affiliation{\addb}
\affiliation{\addc}
\author{Chen Zha}
\affiliation{\adda}
\affiliation{\addb}
\affiliation{\addc}
\author{Yuan Li}
\affiliation{\adda}
\affiliation{\addb}
\affiliation{\addc}
\author{Jin Lin}
\affiliation{\adda}
\affiliation{\addb}
\affiliation{\addc}
\author{Lei Jiang}
\affiliation{\adda}
\affiliation{\addb}
\affiliation{\addc}
\author{Tao Rong}
\affiliation{\adda}
\affiliation{\addb}
\affiliation{\addc}
\author{Tao Jiang}
\affiliation{\adda}
\affiliation{\addb}
\affiliation{\addc}
\author{Shaoyu Zhan}
\affiliation{\adda}
\affiliation{\addb}
\affiliation{\addc}
\author{Xueshen Wang}
\affiliation{\addc}
\author{Shijian Wang}
\affiliation{\addc}
\author{Wenhui Cao}
\affiliation{\addc}
\author{Yeru Wang}
\affiliation{\adde}
\author{Xiaomin Wang}
\affiliation{\adde}

\author{Ming Gong}
\affiliation{\adda}
\affiliation{\addb}
\affiliation{\addc}
\author{Yulin Wu}
\affiliation{\adda}
\affiliation{\addb}
\affiliation{\addc}
\author{Yuhuai Li}
\affiliation{\adda}
\affiliation{\addb}
\affiliation{\addc}
\author{Gang Wu}
\affiliation{\adda}
\affiliation{\addb}
\affiliation{\addc}
\affiliation{\addd}
\author{Hui Deng}
\affiliation{\adda}
\affiliation{\addb}
\affiliation{\addc}
\author{Kai Yan}
\affiliation{\adda}
\affiliation{\addb}
\affiliation{\addc}
\author{Sirui Cao}
\affiliation{\adda}
\affiliation{\addb}
\affiliation{\addc}
\author{Futian Liang}
\affiliation{\adda}
\affiliation{\addb}
\affiliation{\addc}
\author{Fusheng Chen}
\affiliation{\adda}
\affiliation{\addb}
\affiliation{\addc}
\author{Qingling Zhu}
\affiliation{\adda}
\affiliation{\addb}
\affiliation{\addc}
\author{Yangsen Ye}
\affiliation{\adda}
\affiliation{\addb}
\affiliation{\addc}
\author{Hao Rong}
\affiliation{\adda}
\affiliation{\addb}
\affiliation{\addc}
\author{Shaojun Guo}
\affiliation{\adda}
\affiliation{\addb}
\affiliation{\addc}
\author{Haoran Qian}
\affiliation{\adda}
\affiliation{\addb}
\affiliation{\addc}
\author{Haibin Zhang}
\affiliation{\addc}
\author{Yisen Hu}
\affiliation{\adda}
\affiliation{\addb}
\affiliation{\addc}

\author{Shaowei Li}
\thanks{liswer@hfnl.cn}
\affiliation{\adda}
\affiliation{\addb}
\affiliation{\addc}
\author{Jinjin Li}
\thanks{jinjinli@nim.ac.cn}
\affiliation{\addc}
\author{Cheng-Zhi Peng}
\affiliation{\adda}
\affiliation{\addb}
\affiliation{\addc}
\author{Xiaobo Zhu}
\affiliation{\adda}
\affiliation{\addb}
\affiliation{\addc}
\affiliation{\adde}
\author{Jian-Wei Pan}
\affiliation{\adda}
\affiliation{\addb}
\affiliation{\addc}
\newcommand{\adda}{Hefei National Research Center for Physical Sciences at the Microscale and School of Physical Sciences, University of Science and Technology of China, Hefei 230026, China}
\newcommand{\addb}{Shanghai Research Center for Quantum Science and CAS Center for Excellence in Quantum Information and Quantum Physics, University of Science and Technology of China, Shanghai 201315, China}
\newcommand{\addc}{Hefei National Laboratory, University of Science and Technology of China, Hefei 230088, China}
\newcommand{\addd}{University of Science and Technology of China, Shanghai Research Institute, Shanghai 201315, China}
\newcommand{\adde}{Jinan Institute of Quantum Technology and Hefei National Laboratory Jinan Branch, Jinan 250101, China}
\begin{abstract}
    Superconducting quantum computing faces a critical scalability bottleneck arising from the escalating complexity of traditional analog control infrastructure.
    Single-Flux-Quantum~(SFQ) systems operate at cryogenic temperatures and offer a potential route to addressing these challenges. 
    Using microwave signals at qubit-dependent frequencies to trigger SFQ pulses requires many independent analog microwave sources, limiting the potential to reduce control wiring and increase integration density. A scalable SFQ control system therefore needs a scheme that can address qubits with different transition frequencies using a common clock.
    Here, we propose and experimentally demonstrate a control scheme for implementing both single- and two-qubit gates. The scheme uses guiding waveforms to rapidly generate bipolar high-speed pulse sequences on a common clock, providing a basis for future implementation in SFQ control systems.
    Numerically, this scheme achieves a simulated fidelity of 99.99\% for 40~ns $R(\varphi, \frac{\pi}{2})$ gates with arbitrary initial phases and 99.9\% for 100~ns controlled-Z~(CZ) gates. Experimentally, by utilizing an arbitrary waveform generator~(AWG) operated at a fixed sampling rate of 25~GSa/s to deliver the bipolar high-speed pulse sequences, we successfully demonstrate 40~ns single-qubit gates with an average gate fidelity of 99.94\% estimated using cross-entropy benchmarking~(XEB) and 120~ns CZ gates with 99.63\% fidelity. These results demonstrate the feasibility of common-clock encoded pulse control for both single- and two-qubit gates.
\end{abstract}
\maketitle
\section{Introduction}
Quantum computing exploits superposition and entanglement~\cite{steane1998quantum, Nielsen_Chuang_2010, Sakurai_Napolitano_2020} to accelerate selected computational tasks, as exemplified by Shor's and Grover's algorithms~\cite{shor1994algorithms, grover1996fast}. Superconducting circuits are a promising platform owing to their compatibility with established micro- and nanofabrication technologies~\cite{clarke2008superconducting, josephson1962possible, jaklevic1964quantum}. Large-scale superconducting quantum computing requires many qubits, but increasing the qubit count places growing demands on electronic control resources and interconnections to the dilution refrigerator~\cite{jordan2026quantum}. Additional cables also increase the refrigeration load~\cite{krinner2019engineering}. Integrating control electronics into the dilution refrigerator, for example using single-flux-quantum~(SFQ) circuits~\cite{likharev1991rsfq}, offers a potential route to addressing these scaling challenges.

Previous studies~\cite{liu2023single, leonard2019digital, mcdermott2014accurate, PRXQuantum.3.010350, jordan2026quantum} have used SFQ pulse sequences to implement single-qubit gates, with randomized benchmarking~(RB) yielding an average Clifford fidelity of approximately 99.9\%~\cite{liu2026}. In schemes that use equally spaced SFQ pulses, addressing qubits with different transition frequencies simultaneously requires each dc-to-SFQ converter to receive a microwave trigger at a subharmonic $f_q/n$ of its target qubit's transition frequency $f_q$, where $n$ is a positive integer. Providing separate trigger channels for different frequencies limits the potential to reduce wiring and increase control-system integration.
To generate control signals without relying on qubit-specific microwave trigger frequencies, Liebermann and Wilhelm~\cite{liebermann2016optimal} proposed implementing single-qubit gates with high-speed SFQ pulse sequences defined on a fixed clock and used a genetic algorithm to optimize the pulse patterns. Li et al.~\cite{li2019hardware} proposed a common-clock scheme that constructs single-qubit gates by repeating short SFQ pulse subsequences, reducing the amount of sequence data that must be stored. However, these encoding schemes remain experimentally unverified, and genetic-algorithm searches can be computationally demanding. Existing SFQ-based two-qubit gate proposals are limited to architectures without couplers and remain theoretical~\cite{PhysRevApplied.19.044031,jokar2021practical}.
Here, we propose a common-clock control scheme that uses encoded bipolar high-speed pulse sequences to implement both single- and two-qubit gates. We develop a method for rapidly generating these sequences from guiding waveforms and evaluate it in simulations. The use of bipolar rather than unipolar pulses allows higher pulse densities, stronger effective driving, and cleaner spectra at the same sampling rate, enabling higher-fidelity quantum control~\cite{bastrakova2022genetic}.
We then use a high-speed arbitrary waveform generator~(AWG) to produce the sequences on a common clock and demonstrate high-fidelity single- and two-qubit gates in experiments. These results provide an experimental basis for exploring the scheme in future SFQ control systems.

In this paper, we experimentally demonstrate high-fidelity single-qubit $R(\varphi, \frac{\pi}{2})=\exp(\frac{-i\pi}{4}(\cos\varphi\hat{\sigma_{x}}+\sin\varphi\hat{\sigma_{y}}))$ gates and two-qubit controlled-Z (CZ) gates~\cite{li2022realization} using encoded bipolar high-speed pulse sequences on a common clock. A guiding waveform is used to generate the sequences and simplify experimental calibration. Before the pulses are inductively coupled to the transmon system~\cite{schreier2008suppressing}, microwave filters~\cite{pozar2021microwave} attenuate unwanted high-frequency components while preserving the spectral components needed to drive the gates.
Employing an arbitrary waveform generator operated at a fixed sampling rate of 25~GSa/s to synthesize the digital streams and using cross-entropy benchmarking~(XEB)~\cite{boixo2018characterizing}, we obtain an average gate fidelity of 99.94\% for the 40~ns single-qubit $R(\varphi,\pi/2)$ gate and approximately 99.63\% for the 120~ns CZ gate. These results provide an AWG-based experimental validation of the encoded control protocol and a basis for investigating its future implementation with SFQ circuits.
\begin{figure*}[!htbp]
    \centering \resizebox{17cm}{!}{\includegraphics{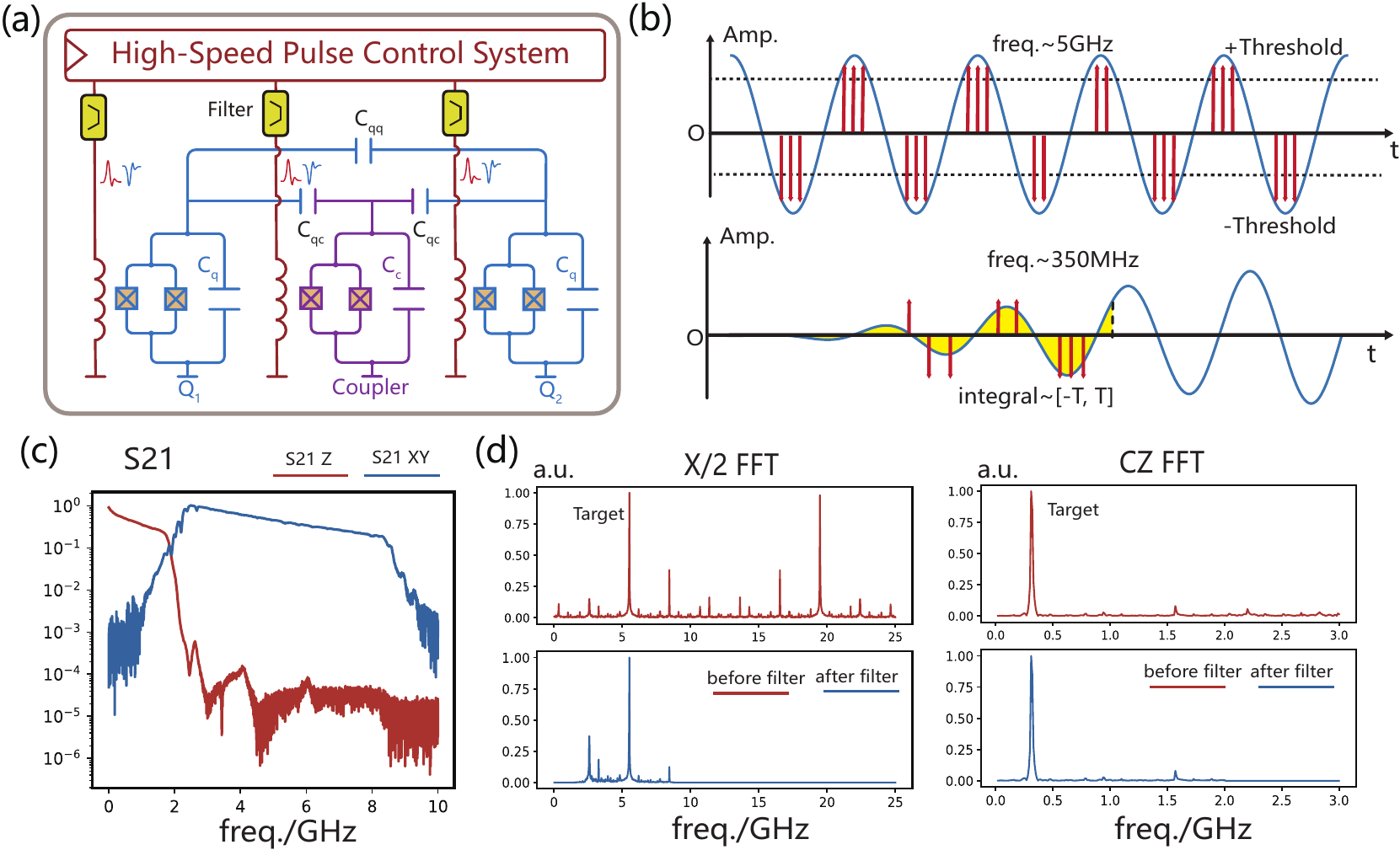}}
    \caption{\textbf{Control Scheme and High-Speed Pulse-Sequence Generation Methods} (a) The circuit diagram of the coupled qubits and the high-speed pulse control system used in this study. (b) Two methods for encoding high-speed pulse sequences based on the guiding waveform (blue). (c) The $S_{21}$ response of the XY line and Z line used in the experiment. The XY line has a 2--8~GHz bandpass filter, while the Z line has a 2~GHz low-pass filter. (d) The fast Fourier transform~(FFT) of the control waveforms for the X/2 gate and CZ gate in the simulation. Red and blue indicate the signals before and after filtering, respectively.}
    \label{fig_one}
\end{figure*}

\section{Device and Scheme} 
Fig.~\ref{fig_one}(a) shows the overall schematic used in this paper. The blue components represent the qubits, while the purple component represents the coupler, which mediates the coupling between adjacent qubits and enables the CZ gate~\cite{yan2018tunable}.

We base our simulation on a model consisting of two qubits for simplicity. The Hamiltonian of the qubit--coupler--qubit model is~\cite{yan2018tunable}
\begin{align}
    \hat{H}_{\text{total}}= & \sum_{i=1,2,c}(-E_{J,i}\cos{(\frac{\phi_{r,i}+\phi_{z,i}(t)}{2})}\cos{(\hat{\varphi}_{i}+\phi_{x,i}(t))}\nonumber \\&+4E_{C,i}\hat{n}_{i}^{2})+\hat{H}_{\text{int}}\\
    \hat{H}_{\text{int}}=   & 8E_{C,12}\hat{n}_{1}\hat{n}_{2}+\sum_{i=1,2}8E_{C,ic}\hat{n}_{i}\hat{n}_{c}.
\end{align}
In this equation, $\hat{H}_{\text{total}}$ is the total Hamiltonian, and $\hat{H}_{\text{int}}$ is the interaction Hamiltonian induced by capacitive coupling between the qubit and coupler. $E_{C}$ and $E_{J}$ are the charging energy and Josephson energy of the qubits and the coupler, respectively. $\phi_{r}=\frac{2\pi}{\Phi_{0}}\Phi_{r}$ is the reduced residual flux which can be controlled through the DC Z channel to adjust the frequencies of the qubits and coupler, mitigating the residual coupling between qubits. $\phi_{z}(t)=\frac{2\pi}{\Phi_{0}}\Phi_{z}$ represents the Z-channel reduced flux signal and $\phi_{x}(t)=\frac{2\pi}{\Phi_{0}}\Phi_{x}$ represents the X-channel reduced flux signal. The indices $i=1,2$ label qubits $Q_{1}$ and $Q_{2}$, respectively, while the index $c$ labels the coupler. In the simulations, $\phi_{x, i}(t)$ and $\phi_{z, i}(t)$ are replaced by high-speed pulse sequences, and we retain four levels for each of $Q_1$, the coupler~(C), and $Q_2$, giving a total Hilbert-space dimension of \(4 \times 4 \times 4 = 64\).
\begin{align}
    \hat{\varphi}=(\frac{8E_{C}}{E_{J}})^{\frac{1}{4}}\frac{\hat{a}^{\dagger}+\hat{a}}{\sqrt{2}}\nonumber \\
    \hat{n}=(\frac{E_{J}}{8E_{C}})^{\frac{1}{4}}\frac{\hat{a}^{\dagger}-\hat{a}}{\sqrt{2}i}.
    \label{eq1}
\end{align}
$\hat{n}$ and $\hat{\varphi}$ are the Cooper-pair number operator and the reduced phase operator, respectively. $\hat{a}$ and $\hat{a}^{\dagger}$ are the annihilation operator and creation operator, respectively.
The presence of the coupler can reduce unwanted residual coupling between adjacent qubits~\cite{yan2018tunable}, increasing the independence between qubits and the fidelity of quantum gates.

The high-speed pulse control system generates encoded bipolar high-speed pulse sequences. In our experiments, these sequences are generated by a high-speed AWG operated at a sampling rate of 25~GSa/s.
The sequences contain multiple frequency components.
Filters are placed between the pulse source and the quantum chip to suppress unwanted components [Fig.~\ref{fig_one}(d)], with their $S_{21}$ responses shown in Fig.~\ref{fig_one}(c).

Fig.~\ref{fig_one}(b) illustrates two methods for generating sequences. When generating pulse sequences, a guiding waveform~($W(t)$) is required, as indicated by the blue curve in the figure.
\begin{equation}
    W(t) = A(t)\sin(\omega t + \varphi_{0}).
\end{equation}
$A(t)$ is the amplitude envelope of the guiding waveform. For single-qubit $R(\varphi,\frac{\pi}{2})$ gates, $A(t)$ is a constant, while for the CZ gate, $A(t)$ is a trapezoidal function.

\begin{algorithm}[H]
\caption{Integration-based encoding of a low-frequency guiding waveform}
\label{alg:integration_encoding}
\begin{algorithmic}[1]
\Require Guiding waveform $W(t)$; clock interval $\Delta t$;
         number of slots $N$; threshold $T>0$;
         pulse-area magnitude $S_{\mathrm{p}}>0$
\Ensure Bipolar pulse sequence $\{s_k\}_{k=0}^{N-1}$
\State $r \gets 0$ \Comment{Accumulated residual area}
\For{$k=0,\ldots,N-1$}
    \State $r \gets r +
        \displaystyle\int_{k\Delta t}^{(k+1)\Delta t} W(t)\,dt$
    \If{$r \geq T$}
        \State $s_k \gets +1$
        \State $r \gets r-S_{\mathrm{p}}$
    \ElsIf{$r \leq -T$}
        \State $s_k \gets -1$
        \State $r \gets r+S_{\mathrm{p}}$
    \Else
        \State $s_k \gets 0$
    \EndIf
\EndFor
\State \Return $\{s_k\}_{k=0}^{N-1}$
\end{algorithmic}
\end{algorithm}

\begin{algorithm}[H]
\caption{Threshold-based encoding of a high-frequency guiding waveform}
\label{alg:threshold_encoding}
\begin{algorithmic}[1]
\Require Guiding waveform $W(t)$; clock interval $\Delta t$;
         number of slots $N$; positive threshold $T_{+}>0$;
         negative threshold $T_{-}<0$
\Ensure Bipolar pulse sequence $\{s_k\}_{k=0}^{N-1}$
\For{$k=0,\ldots,N-1$}
    \State $w_k \gets W(k\Delta t)$
    \If{$w_k > T_{+}$}
        \State $s_k \gets +1$
    \ElsIf{$w_k < T_{-}$}
        \State $s_k \gets -1$
    \Else
        \State $s_k \gets 0$
    \EndIf
\EndFor
\State \Return $\{s_k\}_{k=0}^{N-1}$
\end{algorithmic}
\end{algorithm}

During the process of generating high-speed pulse sequences, time is divided into many small intervals, each of which has a duration of approximately 40~ps~\cite{li2019hardware,liebermann2016optimal,bastrakova2022genetic}. Generating high-speed pulse sequences involves determining whether each small time interval should be encoded as 0, 1, or -1.
For low-frequency signals, such as those near the transition frequency between $\ket{11}$ and $\ket{02}$, we integrate the guiding waveform starting at $t=0$. When the accumulated value reaches the positive or negative threshold, we generate a pulse of the corresponding polarity and subtract its signed area from the accumulated value, retaining the residual for subsequent time steps. This method encodes the guiding waveform through the local pulse density while tracking its accumulated area. When the clock frequency is much higher than the target frequency, this approach can better preserve the waveform after low-pass filtering than direct thresholding, which can introduce low-order harmonics within the filter passband.
For high-frequency signals, typically around 5~GHz and close to the qubit transition frequency $f_{01}$, we use positive and negative thresholds. At each clock slot, a positive pulse (+1) is generated if the guiding waveform exceeds the positive threshold, and a negative pulse (-1) is generated if it falls below the negative threshold. Otherwise, no pulse (0) is generated. At a clock frequency of 25~GHz, a 5~GHz waveform has only about five clock slots per period, leaving limited scope for encoding its local amplitude through pulse density. Direct thresholding instead determines pulse placement from the instantaneous guiding waveform, providing a simple way to generate the near-resonant drive. We therefore use thresholding in this frequency range rather than the integration-based method.
\begin{figure*}[!htbp]
    \centering \resizebox{17cm}{!}{\includegraphics{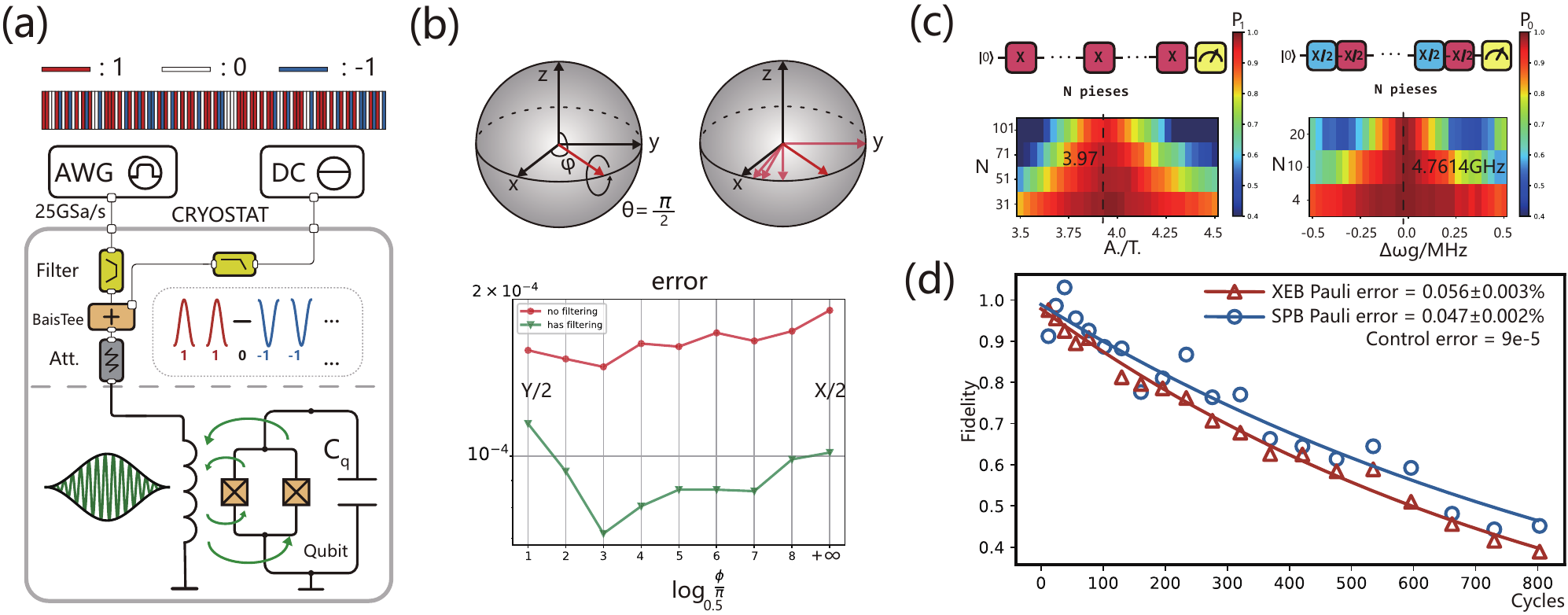}}
    \caption{\textbf{Single-Qubit Gates: Simulation and Experimental Results} (a) The experimental setup and schematic for implementing single-qubit $R(\varphi,\frac{\pi}{2})$ gates using an AWG to generate high-speed pulse sequences. The AWG signal~(providing pulses) and DC line~(providing the qubit's Z-bias) are combined in the dilution refrigerator using a bias tee. Before the merged signal interacts with the qubit, it passes through attenuators and filters. The sequence of high-speed pulses is composed of $+1$~(red), $-1$~(blue), and $0$~(white) elements. (b) The representation of the rotation axis and rotation angle for single-qubit operations, where $\varphi$ and $\theta$ are the azimuthal and polar angles of the rotation axis, respectively, and $\phi$ is the rotation angle. In the simulation, the shape of high-speed pulses is not important. The bottom panel illustrates the fidelity of $R(\varphi, \frac{\pi}{2})$ gates under different phases of the guiding waveforms. The red curve represents the scenario without filters, while the green curve represents the scenario with filters. (c) The calibration process of single-qubit $R(\varphi,\frac{\pi}{2})$ gates. The left panel shows an odd number of high-speed-pulse-based X gates together, and the final measurement of the probability of occupying state $\ket{1}$ determines the amplitude-to-threshold ratio of the guiding waveform~(A./T.). The right panel shows an even number of $(X/2,-X/2)$ gate pairs applied consecutively. The final population in state $\ket{0}$ is measured to determine the guiding-waveform frequency. (d) XEB results for single-qubit $R(\varphi,\frac{\pi}{2})$ gates, where the red curve represents the XEB curve, and the blue curve represents the speckle purity benchmarking~(SPB)~\cite{arute2019quantum} curve. In this experiment, the XEB-derived average gate fidelity is $(99.940\pm0.003)\%$ and the control error is $9\times10^{-5}$.}
    \label{fig_two}
\end{figure*}

\section{Single-Qubit Gates}
\textit{Simulation.}
As shown in Fig.~\ref{fig_two}(a), due to the presence of the coupler, a transmon qubit can be considered separately.
Our pulse sequences are generated from a sinusoidal guiding waveform. There are three key parameters: frequency, threshold, and initial phase. The guiding-waveform frequency matches the qubit transition frequency $f_{01}$, the threshold controls the effective drive strength, and the initial phase determines the azimuthal angle of the gate's rotation axis on the Bloch sphere ($\varphi$ in Fig.~\ref{fig_two}(b)).

These single-qubit gates can achieve high fidelity if other high-frequency peaks in the spectrum do not coincide with other transition frequencies and leakage~\cite{krantz2019quantum} from $\ket{1}$ to $\ket{2}$ is suppressed. Approximately, the leakage $|\bra{2}\hat{U}(t)\ket{1}|^{2}$ will oscillate at a frequency of $\eta$, which is	the qubit anharmonicity, so to suppress the leakage the gate time can be chosen as
\begin{equation}
    T_{\mathrm{g}} = \frac{2n\pi}{|\eta|}\quad n=1,2,3,\ldots.
\end{equation}
In the simulation, the qubit's $f_{01}$ is $5.5175~\text{GHz}$ and the anharmonicity $\eta$ is $-2\pi\times 224.635~\text{MHz}$. $T_{\mathrm{g}}$ was chosen as $40~\text{ns}$, so the leakage probability undergoes approximately nine oscillation cycles and returns close to zero.
The threshold controls the pulse-sequence duty cycle and thus the Rabi oscillation frequency, allowing the effective drive strength to be adjusted even when the individual pulse amplitude is fixed.

In our simulations, we evaluate the average fidelity of a single-qubit $R(\varphi,\pi/2)$ gate by averaging over the six cardinal states of the Bloch sphere:
\begin{equation}
    \label{fidelity}
    F_{\mathrm{avg}} =
    \frac{1}{6}\sum_{\alpha\in\mathcal{V}}
    \left|
    \bra{\alpha}
    R^{\dagger}(\varphi,\pi/2)M
    \ket{\alpha}
    \right|^{2},
\end{equation}
where
\begin{equation}
    \mathcal{V} =
    \left\{
    \ket{0},\ket{1},
    \frac{\ket{0}\pm\ket{1}}{\sqrt{2}},
    \frac{\ket{0}\pm i\ket{1}}{\sqrt{2}}
    \right\},
\end{equation}
and
\begin{equation}
    R(\varphi,\pi/2) =
    \exp\left[
    -\frac{i\pi}{4}
    \left(\cos\varphi\,\hat{\sigma}_x
    +\sin\varphi\,\hat{\sigma}_y\right)
    \right]
\end{equation}
is the ideal target gate. To obtain $M$, we first project the full 64-dimensional evolution operator onto the dressed two-qubit computational subspace, ordered as $\{\ket{00},\ket{01},\ket{10},\ket{11}\}$. The coupler is assumed to remain in its ground state throughout the single-qubit gate operation, and its state label is omitted. We then extract the $2\times2$ block associated with $\{\ket{00},\ket{10}\}$, corresponding to operations on the first qubit with the second qubit initially and finally in $\ket{0}$. The projected output states are not renormalized, so population transferred outside this selected subspace reduces the reported fidelity.

Using our simulation parameters, we generated a set of single-qubit $R(\varphi,\pi/2)$ gates and compared them with the ideal operations using Eq.~(\ref{fidelity}). The resulting errors are shown by the red curve in Fig.~\ref{fig_two}(b). All simulated gates achieve fidelities of at least 99.98\%.

To reduce the impact of unwanted high-frequency components on the qubits, our scheme includes attenuators and filters between the high-speed pulse generators and the qubits. Accordingly, we include a 2--8~GHz bandpass filter in the simulations.
The errors of these single-qubit $R(\varphi,\frac{\pi}{2})$ gates are shown as the green curve in Fig.~\ref{fig_two}(b).
It is worth noting that with the addition of filtering, the fidelities of most gates are around 99.99\%, which are higher than those obtained without filtering.

\textit{Experiment.}
In the experiment, we used a Tektronix AWG70002A to generate the high-speed pulse sequences. This type of AWG can reach a sampling rate of 25~GSa/s, and can generate positive and negative pulses, with a peak-to-peak voltage~($V_{\mathrm{pp}}$) of up to 500~mV. Fig.~\ref{fig_two}(a) is a schematic diagram of our experimental equipment. The pulse sequence is used to implement quantum gate operations, and the DC signal is used to modulate the qubit's frequency. The two signals are combined by an adder in the refrigerator and added to the superconducting rings of the qubit through mutual inductance. The XY line includes a 2--8~GHz bandpass filter.

Initially, we used classical electronics to calibrate the basic parameters of the qubit. The qubit frequency was set to 4.76~GHz and its anharmonicity was $2\pi\times 240~\text{MHz}$. The calibrated qubit frequency was used as the initial
frequency of the pulse sequence's guiding waveform. In the experiment, the gate time was also set to 40~ns. According to the sampling frequency of the AWG, there were 1000 high-speed pulse slots to be determined.

We set the AWG sampling rate to 25~GSa/s and fixed the AWG output amplitude to 160~mV (the amplitude of the high-speed pulse cannot be adjusted). In each 40~ps time slot, the AWG output takes one of three values: the maximum voltage, the minimum voltage, or zero.

After calibrating the frequency and threshold of the guiding waveform~(Fig.~\ref{fig_two}(c)), we set the gate phase through the guiding-waveform phase according to
\begin{equation}
    \varphi_{\text{0}} = 2\pi f_{\text{guiding}}t_{\text{start}}+ \varphi,
\end{equation}
where $f_{\text{guiding}}$ and $\varphi_{\text{0}}$ are the frequency and phase of the guiding waveform, respectively. $t_{\text{start}}$ is the start point of the gate in the quantum circuit. $\varphi$ denotes the azimuthal angle of the single-qubit gate's rotation axis on the Bloch sphere.

After achieving precise control over the phases of single-qubit $R(\varphi,\frac{\pi}{2})$ gates, we used XEB to characterize their fidelity. We varied the phase of the guiding waveform to generate the corresponding high-speed pulse sequences, thereby randomly generating $R(\varphi,\frac{\pi}{2})$ with different phases. We then compared the experimental values with the theoretical values. The XEB-derived average gate infidelity is $(0.0562\pm0.0035)\%$, corresponding to an average gate fidelity of 99.94\%~[Fig.~\ref{fig_two}(d)]. The uncertainty is a standard-error estimate propagated from the fitted decay parameter. Meanwhile, the control error is below $10^{-4}$, consistent with the simulation results.
\begin{figure*}[!htbp]
    \centering \resizebox{17cm}{!}{\includegraphics{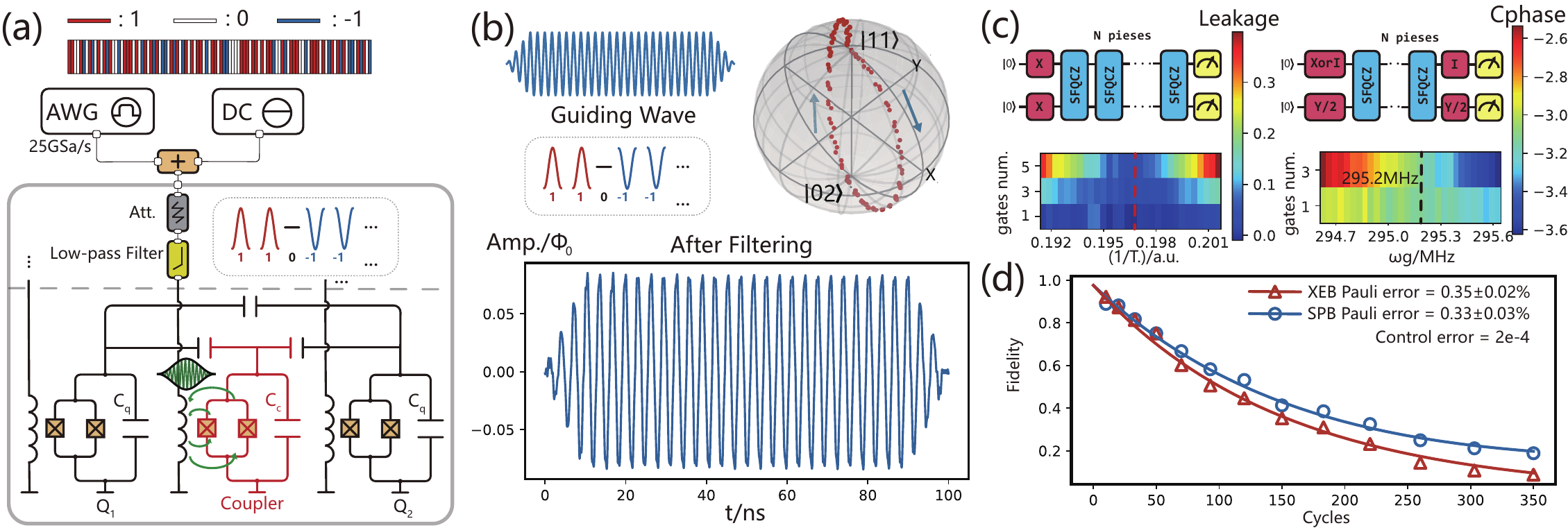}}
    \caption{\textbf{CZ Gate: Simulation and Experimental Results} (a) The experimental schematic for the CZ gate. The high-speed AWG~(generating high-speed pulse sequences) and the DC line~(setting the coupler's cutoff point) are combined at room temperature using an adder onto a single Z-line, which is then introduced into the dilution refrigerator. Before the signal reaches the coupler, it passes through filters and attenuators. (b) The upper-left panel is a schematic of the guiding waveform. The lower-left panel shows the signal after the pulse sequence has been filtered~(closely resembling the guiding waveform). The upper-right panel depicts the simulated evolution of the pulse-sequence-based CZ gate~(on the Bloch sphere of the $\ket{11}$ and $\ket{02}$ states). The Bloch vector, initially in the $\ket{11}$ state, completes a full rotation on this Bloch sphere within 100~ns. (c) Calibration procedures of the CZ gate: the left panel shows multiple CZ gates applied consecutively. By scanning the threshold of the guiding waveform~(1/T.), the final threshold is determined based on the leakage probability out of the computational space. The right panel shows a scan of the frequency of the guiding waveform~($\omega_{g}$), calibrated using the conditional phase. (d) XEB results for CZ gates, where the red curve represents the XEB curve, and the blue curve represents the SPB curve. The curves show the cycle-dependent XEB and SPB results. After subtracting the single-qubit Pauli-error contributions to first order, the XEB analysis gives an estimated average CZ gate fidelity of $(99.63\pm0.02)\%$.}
    \label{fig_three}
\end{figure*}

\section{CZ Gate}
\textit{Simulation.}
As shown in Fig.~\ref{fig_three}(a), the CZ pulse sequence is optimized in advance. The encoded bipolar pulse sequences are also generated using a low-frequency guiding waveform. Compared to the method of generating pulse sequences using thresholds, the signals generated using the integration method better preserve the shape of the guiding waveform after passing through a filter. The filter not only relaxes the amplitude constraint of the high-speed pulses before filtering~(excessive high-speed pulse amplitude may cause the coupler's Z-bias to exceed the half-flux point), but also improves the fidelity of the pulse-sequence-based CZ gate compared to the case without filters.

First, we calibrate the corresponding microwave signal capable of implementing a CZ gate with fidelity exceeding $99.9\%$ by sweeping the amplitude and frequency of the signal applied to the coupler~\cite{Li_2022}. Then, we utilize this microwave signal as the guiding waveform for the pulse sequence. In the simulation, the frequency difference between the energy levels $\ket{11}$ and $\ket{02}$ is found to be 315.7~MHz, while the calibrated guiding-waveform frequency is 314.5~MHz.

We evaluate the average CZ gate fidelity as~\cite{pedersen2007fidelity}
\begin{equation}
    \label{eq:cz_fidelity}
    F_{\mathrm{avg}}^{\mathrm{CZ}} =
    \frac{
    \operatorname{Tr}(U_{\mathrm{comp}}^\dagger U_{\mathrm{comp}})
    +
    \left|\operatorname{Tr}
    (U_{\mathrm{CZ}}^\dagger U_{\mathrm{comp}})\right|^2
    }{20},
\end{equation}
where $U_{\mathrm{comp}}$ is the $4\times4$ evolution matrix
projected onto the dressed computational subspace
$\{\ket{00},\ket{01},\ket{10},\ket{11}\}$ after local single-qubit
phase compensation, and
$U_{\mathrm{CZ}}=\operatorname{diag}(1,1,1,-1)$ is the ideal CZ gate.
The projected matrix is not renormalized, so leakage reduces
the reported fidelity.

Using Eq.~(\ref{eq:cz_fidelity}), we obtain a simulated CZ gate
fidelity of 99.9\% at a pulse-sequence sampling rate of 25~GSa/s,
as shown in Fig.~\ref{fig_three}(b).

\textit{Experiment.}
Similarly, we used the AWG70002A to generate high-speed pulse sequences. The AWG was operated as a bipolar DAC at a sampling rate of 25~GSa/s, with a maximum output amplitude of 200~mV.

The experimental setup is depicted in Fig.~\ref{fig_three}(a). At room temperature, the AWG and DC line are combined through an adder into a single Z-line, which is then introduced into the dilution refrigerator. The DC line provides a DC signal to set the coupler's cutoff point. The AWG generates pulse sequences to drive the transition between $\ket{11}$ and $\ket{02}$. On the Z-line, there are low-pass filters to remove the high-frequency components of the high-speed pulses, reducing the impact of unwanted spectral components on the quantum system.

The frequency difference between the levels $\ket{11}$ and $\ket{02}$ was calibrated using traditional electronics. This frequency served as the initial frequency of the guiding waveform. Next, we calibrated the guiding-waveform threshold by minimizing leakage from the computational subspace. We calibrated the guiding-waveform frequency by measuring the conditional phase [Fig.~\ref{fig_three}(c)]. Note that for achieving high-fidelity gates, frequency and threshold were calibrated iteratively. Ultimately, we obtained the calibrated guiding waveform and the corresponding high-speed pulse sequence for the CZ gate. 

We characterized the pulse-sequence-based CZ gate using XEB. Each cycle consisted of simultaneous single-qubit gates on both qubits followed by one CZ gate. The single-qubit gates were implemented using conventional control electronics, while the AWG generated the CZ pulse sequences. Under an independent, small-error approximation, we estimated the CZ Pauli error as~\cite{arute2019quantum}
\begin{equation}
    e_{\mathrm{P,CZ}}
    \approx
    \frac{15}{16}(1-p_{\mathrm{XEB}})
    -e_{\mathrm{P},1}
    -e_{\mathrm{P},2},
\end{equation}
where $p_{\mathrm{XEB}}$ is the fitted cycle-decay parameter and $e_{\mathrm{P},1}$ and $e_{\mathrm{P},2}$ are the independently characterized Pauli errors of the single-qubit gates used in the experiment. The estimated average CZ gate fidelity is
\begin{equation}
    F_{\mathrm{avg}}^{\mathrm{CZ}}
    \approx 1-\frac{4}{5}e_{\mathrm{P,CZ}}.
\end{equation}
For the 120~ns CZ gate, this analysis gives a Pauli error of $(0.46\pm0.02)\%$ and an estimated average gate fidelity of $(99.63\pm0.02)\%$. The quoted uncertainties are standard-error estimates propagated from the fitted cycle-decay parameter and do not include uncertainties in the single-qubit reference errors. The control error of the CZ gate is $2\times10^{-4}$.
\section{Conclusion}
This paper proposes a scheme to control superconducting quantum circuits with couplers using encoded bipolar high-speed pulse sequences.
Based on our proposed scheme, different quantum gates use the same trigger frequency, which simplifies the trigger system.
All the high-speed pulse sequences are generated using the guiding-waveform method, which saves generation time in experiments compared to genetic algorithm optimization.
Additionally, the high-speed pulse sequences undergo filtering to eliminate high-frequency signals that can cause unwanted leakage, thereby significantly enhancing the fidelity of the quantum gate. In our experiment, we use the high-speed AWG to generate high-speed pulse sequences and achieve single-qubit $R(\varphi,\frac{\pi}{2})$ gates with an XEB-derived average gate fidelity exceeding 99.94\% and CZ gates with an XEB-derived average gate fidelity of approximately 99.63\%. Together, the single-qubit rotations and the CZ gate constitute a universal gate set.

The digital control scheme for qubits presented in this paper can be applied to future SFQ systems which are based on high-speed SFQ pulses.
In future work, we plan to exploit the low power dissipation of superconducting circuits by integrating SFQ waveform generators with qubits so that both operate at 20~mK. The encoded SFQ pulse sequences would be stored in superconducting digital circuitry and selected using digital commands sent over a small number of external control lines. A common clock line would time the playback of these sequences to implement the universal gate set required for scalable quantum computing.

The authors thank the USTC Center for Micro- and Nanoscale Research and Fabrication for supporting the sample fabrication and QuantumCTek Co., Ltd. for supporting the fabrication and maintenance of room-temperature electronics. 
This research was supported by the Quantum Science and Technology-National Science and Technology Major Project (Grant No. 2021ZD0300200), the Cultivation Project of Shanghai Research Center for Quantum Sciences (Grant No. LZPY2024), the Shanghai Municipal Science and Technology Major Project (Grant No. 2019SHZDZX01), and the National Natural Science Foundation of China (Grant Nos. 92476203 and 92476001), as well as special funds from the Jinan Science and Technology Bureau and the Jinan High-tech Zone Management Committee. Y.X. acknowledges support from the Shanghai Sailing Program. Y.S. Ye acknowledges support from the National Natural Science Foundation of China (Grant No. 12404575), the Key R\&D Plan of Shandong Province (Grant No. 2024CXPT083), and the National Key Research and Development Program of China (Grant No. 2024YFB4504002). X.B. Zhu acknowledges support from the Taishan Scholars Program and the New Cornerstone Science Foundation through the XPLORER PRIZE.
\bibliography{main.bib}
\end{document}